\documentclass[twocolumn,color,superscriptaddress,psfig,showpacs,amsmath,amssymb,floatfix,fleqn]{revtex4-1}
\usepackage{epsf}

\usepackage{amssymb}

\usepackage{natbib}
\setcitestyle{square,numbers}

\usepackage{graphicx}
\usepackage{dcolumn}
\usepackage{bm}
\usepackage{color}

\usepackage{epstopdf}

\newcommand{\BT}{Bi$_2$Te$_3$}
\newcommand{\BTM}{Bi$_{\rm 2-x}$Mn$_{\rm x}$Te$_3$}
\newcommand{\BS}{Bi$_2$Se$_3$}
\newcommand{\BSM}{Bi$_{\rm 2-x}$Mn$_{\rm x}$Se$_3$}
\newcommand{\ST}{Sb$_2$Te$_3$}

\begin{document}


\title{Spin dynamics and magnetic interactions of Mn dopants \\ in the topological insulator Bi$_2$Te$_3$}

\author{S.\ Zimmermann}
 \affiliation{Leibniz Institute for Solid State and Materials Research IFW Dresden, 01171 Dresden, Germany}
 \affiliation{Institute for Solid State Physics, TU Dresden, 01069 Dresden, Germany}

\author{F.\ Steckel}%
 \affiliation{Leibniz Institute for Solid State and Materials Research IFW Dresden, 01171 Dresden, Germany}
 \affiliation{Center for Transport and Devices, TU Dresden, 01069 Dresden, Germany}

\author{C.\ Hess}
 \affiliation{Leibniz Institute for Solid State and Materials Research IFW Dresden, 01171 Dresden, Germany}
 \affiliation{Center for Transport and Devices, TU Dresden, 01069 Dresden, Germany}

\author{H.\ W.\ Ji}
 \affiliation{Department of Chemistry, Princeton University, 08544 Princeton, New Jersey, USA}

\author{Y.\ S. Hor}
 \affiliation{Department of Chemistry, Princeton University, 08544 Princeton, New Jersey, USA}

\author{R.\ J.\ Cava}
 \affiliation{Department of Chemistry, Princeton University, 08544 Princeton, New Jersey, USA}

\author{B.\ B\"{u}chner}%
 \affiliation{Leibniz Institute for Solid State and Materials Research IFW Dresden, 01171 Dresden, Germany}
 \affiliation{Center for Transport and Devices, TU Dresden, 01069 Dresden, Germany}

\author{V.\ Kataev}
 \affiliation{Leibniz Institute for Solid State and Materials Research IFW Dresden, 01171 Dresden, Germany}

\date{\today}

\begin{abstract}

The magnetic and electronic properties of the magnetically doped topological insulator \BTM\, were studied using electron spin resonance (ESR) and
measurements of static magnetization and electrical transport. The investigated high-quality single crystals of \BTM\, show a ferromagnetic
phase transition for $x\geq 0.04$ at $T_{\rm C}\approx 12$\,K. The Hall measurements reveal a $p$-type finite charge-carrier density. Measurements
of the temperature dependence of the ESR signal of Mn dopants for different orientations of the external magnetic field give evidence that the
localized Mn moments interact with the mobile charge carriers leading to Ruderman-Kittel-Kasuya-Yoshida-type ferromagnetic coupling between the Mn spins of order
$2-3$\,meV. Furthermore, ESR reveals a low-dimensional character of  magnetic correlations that persist far above the ferromagnetic ordering
temperature.

\end{abstract}


\maketitle

\section{\label{Intro}Introduction}
Topological insulators (TIs) are currently attracting a great deal of interest in the area of condensed-matter research. This interest is due to their unique surface states, which have a linear band
dispersion (Dirac cone) and a momentum-locked spin of the charge carriers \cite{Hasan2010,Hasan2011,Qi2011}. In such insulating materials, a strong
spin-orbit coupling plays a key role. It yields a band inversion in the bulk, whereas the valence and conduction bands ultimately cross at the Dirac point at the material's surface. This gives rise to the gapless conducting surface states that are protected from localization by topology as long as the time-reversal symmetry (TRS) is preserved. However, a magnetic field breaking the TRS can open a gap at the Dirac point. It has been found that this
scenario is especially interesting because different phenomena, such as the topological magneto-electric effect \cite{Qi2008}, the quantized anomalous
Hall effect \cite{Yu2010,Chang2013}, the giant magneto-optical effect \cite{Tse2010}, or  magnetic monopoles \cite{Qi2009}, can occur in TIs.

One way to introduce a magnetic field is to dope a TI with magnetic ions, such as, e.g., Mn, in a concentration sufficient to cause a magnetic order of
the dopants and hence an internal magnetic field. In this context, several theoretical works have addressed the magnetic interactions between Mn
dopants in the three-dimensional (3D) TIs \BS , \BT , and \ST\ \cite{Larson2008,Niu2011,Henk2012,Zhang2013,Li2014,Vergniory2014}. These materials
appear to be advantageous for experiments since they have a relatively large band gap in the bulk and they grow as stoichiometric crystals
\cite{Hasan2011}, though residual bulk conductivity due to crystal defects \cite{Zhang2013} often complicates probing the surface states.
\\
\citet{Larson2008} have studied the entire series of $3d$ transition-metal (TM) dopants in \BT , \BS , and \ST\, within the local-spin-density
approximation (LSDA). Due to the large size mismatch, they predicted no hybridization of the $3d$ orbitals for the early TM ions. Beyond half-filling
of the $3d$ shell, stronger hybridized covalent bonded states were expected. In a later work, \citet{Niu2011} found by first-principles calculations
that the $3d$ orbitals of the Mn$^{3+}$ dopants ($3d^4$) in \BT\, show a strong hybridization with the $5p$ orbitals of the surrounding Te ions. In combination with the octahedral coordination, this produces a strong crystal-field splitting resulting in a high spin $t^{3}_{\rm 2g}e^{1}_{\rm g}$ configuration. The coupling of the Mn impurities was explained by a superexchange mechanism via the Te ions. In contrast, \citet{Zhang2013} predicted a
valence state of $2+$ for Mn in \BT , in agreement with experimental findings \cite{Hor2010,Vobornik2011,Vobornik2014}. This was supported by a
recent work by \citet{Li2014} in which strong indications for a half-filled $3d^5$ configuration (Mn$^{2+}$) with an atomic like high spin state (Hund's
rules) were found. Again, due to the large size mismatch, the $3d$ orbitals do not hybridize with the Te $5p$ orbitals and thus no dominant crystal-field splitting is predicted. The ferromagnetic coupling of the Mn dopants is explained by the weak but long-range Ruderman-Kittel-Kasuya-Yosida (RKKY)
interaction. Also recently \citet{Vergniory2014} investigated the series of $3d$ TM dopants in \BT, \BS\, and \ST\, by first-principles calculations.
Strong hybridization of the $3d$ orbitals with the host is taken into account, and the exchange integrals $J$ between the different neighbors are
calculated. In their findings, the coupling $J$ between the atomic layers is of the double exchange type, and within a layer the exchange is mediated
by free carriers. For thin films where the bulk conductivity is reduced or extinguished, other mechanisms were proposed. Thus, it has been predicted that the
magnetic coupling is mediated by the surface states \cite{Liu2009} or by an enhanced Van Vleck mechanism \cite{Yu2010}.

Electron spin resonance (ESR) spectroscopy is a valuable tool to probe the spin dynamics and interactions of localized moments dissolved in a
nonmagnetic conducting matrix \cite{Barnes1981}. With this aim, ESR studies were previously reported for \BT\, \cite{elKholdi1994,Isber1995} and
\BS\, \cite{Gratens1997} doped with Gd. Von Bardeleben {\it et al.} \cite{Bardeleben2013} investigated Mn-doped thin films of \BS\, in the
ferromagnetic regime ($T\le 6$\,K) by means of ferromagnetic resonance. Another paper \cite{Silva2015} reported ESR studies on \BTM\,
nanocrystals that are embedded in a glass matrix.

In the present work, to obtain experimental insights into the mechanism of magnetic interactions of TM dopants in a 3D topological insulator, we have
systematically investigated with ESR spectroscopy a series of high-quality single crystals of \BTM\, with doping concentrations $x=0.01$, $0.02$,
$0.04$, $0.07$ and $0.09$. The ESR study was complemented by measurements of magnetization and electrical transport. A combined analysis of
the experimental data enables us to conclude that the coupling between the Mn spins that gives rise to ferromagnetic order in \BTM\, with $x\ge 0.04$
at $T_{\rm C} \approx 9-12$\,K occurs via the RKKY mechanism involving as mediators the mobile charge carriers. Interestingly, ESR measurements reveal that
short-range ferromagnetic correlations are still present at temperatures substantially larger than $T_{\rm C}$, suggestive of a predominantly
two-dimensional character of magnetic exchange associated with the layered structure of \BTM.

The paper is organized as follows. In Sect.~\ref{Details}, sample particulars and information on experimental methods are given. Experimental
magnetization, electrical transport, and ESR data are described and analyzed in Sect.~\ref{Results}. A common discussion of the results obtained is
presented in Sect.~\ref{Disc}, followed by the main conclusions in Sect.~\ref{Concl}.

\section{\label{Details}Experimental details}
\subsection{Samples}

The tetradymite crystal structure (space group $R\bar{3}m$) of \BT\, has a rhombohedral lattice where five atomic planes form  quintuple layers
Te-Bi-Te-Bi-Te stacked along the $c$ axis.  Within the quintuple layers, the bonding is much stronger than the van der Waals-type connection in
between the layers. Thus the crystals naturally cleave parallel to the $ab$ basal plane. The coordination of Bi$^{3+}$ by Te$^{2-}$ ions is roughly
octahedral with a small trigonal distortion.

The series of high-quality single crystals of \BTM\, with doping concentrations of $x=0.01$, $0.02$, $0.04$, $0.07$ and $0.09$ investigated in this
work was grown by a modified Bridgeman process from high purity elemental Bi (99.999\,\%), Mn (99.99\,\%) and Te (99.999\,\%), as explained in detail
by \citet{Hor2010}. They were thoroughly characterized by different physical methods in Ref.~\cite{Hor2010}. In particular, it was
shown that the Mn dopants occupy the Bi sites and that they are homogeneously distributed and not clustered, i.e., the grown single crystals are true
dilute magnetic semiconductors. For doping levels of $x\ge 0.04$, a transition to a ferromagnetic state with $T_{\rm C} \approx 9-12$\,K and an easy-axis parallel to the $c$ axis was observed. The electronic transport measurements revealed $p$-type charge carriers. Furthermore, samples from the
same batch have been investigated with x-ray absorption spectroscopy (XAS) and photoemission spectroscopy (PES) \cite{Vobornik2011,Vobornik2014}. It
was shown that Mn possesses a $3d^5$ configuration as a ground state. The $3d^5$ configuration corresponds to a Mn$^{2+}$ valence state and is
consistent with the known acceptor behavior, when substituting Bi$^{3+}$ \cite{Hor2010}.

\subsection{Experimental methods}

The measurements of the magnetization were performed with a superconducting quantum interference device from Quantum Design equipped with a vibrating
sample magnetometer (VSM-SQUID). The available magnetic fields amount to 7\,T and the temperature can be set between $1.8$\,K and $300$\,K. The
external magnetic field was always applied perpendicular to the crystallographic $c$ axis of the studied sample.
\\
For the electrical transport studies, the samples were mounted in a homemade probe head and contacted with silver paste in a standard four- or
six-point configuration. The current was always applied in the basal $ab$ plane in the $x$ direction, and the voltage was measured either in the same
direction or in the perpendicular $y$ direction to obtain the resistivities $\rho_{\rm xx}$ and $\rho_{\rm xy}$, respectively. The probe head was inserted in
a magnetocryostat from Oxford Instruments with a maximal field of 15\,T and the available temperature range $4.2-300$\,K. The external magnetic field
was applied parallel to the crystallographic $c$ axis of the sample, which is denoted as $z$ axis.
\\
ESR was measured with a commercial X-band spectrometer from Bruker with a rectangular resonator working in the TE$_{103}$ mode at the microwave
frequency $\nu \simeq 9.57$\,GHz. The magnetic field could be swept from 0 to 0.9\,T. The signals were detected with the lock-in technique, for which
the external field was modulated by a small $ac$ field with an amplitude of 0.8\,mT at a frequency of 100\,kHz. As a result, the measured ESR signal
is the field derivative of the absorbed microwave power $\mathrm{d}P(H)/\mathrm{d}H$. The samples were placed in a helium gas-flow cryostat from
Oxford Instruments that enabled temperature dependent measurements between $3.7$\,K and $300$\,K. The cryostat was equipped with a goniometer for
rotating the sample with respect to the external magnetic field. To increase the ESR signal intensity, several cleaved pieces of the crystal were
measured simultaneously by placing them on a substrate with the crystallographic $c$ axis perpendicular to the substrate plane. The alignment of the
$a$ and $b$ axes was not attempted since only minor anisotropy effects within the $ab$ plane are expected.
\section{\label{Results}Experimental results and analysis}
\subsection{\label{Mag}Magnetization}
\begin{figure}
\includegraphics[width=0.94\columnwidth]{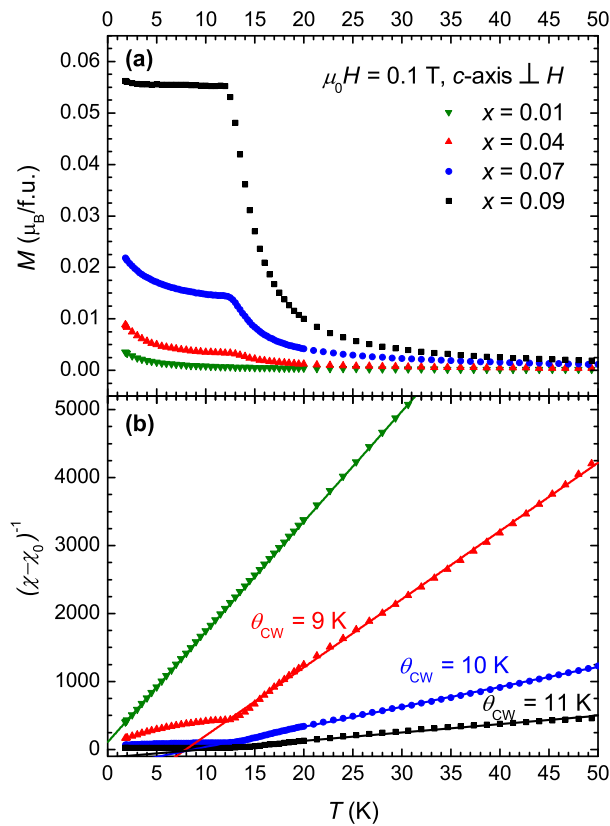}
\caption{Temperature dependence of (a) the static magnetization $M(T)$ and (b) the inverse magnetic susceptibility $\chi(T)^{-1}$ of \BTM, measured
in a weak external field $H$ for various doping concentrations $x$. The external magnetic field $\mu_0H=0.1$\,T was applied perpendicular to the
crystallographic $c$ axis. The solid lines correspond to Curie-Weiss\ fit functions of the linear high temperature regime
$(\chi-\chi_0)^{-1}=(T-\theta_{\rm CW})/C$. (see Sect.~\ref{Mag})} \label{MT}
\end{figure}
The temperature dependence of the magnetization $M(T)$ measured for various doping concentrations ($x=0.01, 0.04, 0.07, 0.09$) is shown in
Fig.~\ref{MT}(a). During the measurements, a weak external magnetic field of $\mu_0H=0.1$\,T was applied perpendicular to the $c$ axis. This field is
in the range where $M$ depends linearly on $H$. For doping concentrations $x\geq 0.04$, a transition to the ferromagnetic regime at $T\approx 12$\,K
is visible as a kink in the $M(T)$ curve. A precise determination of the transition temperature $T_{\rm C}$ has already been done by
\citet{Hor2010} using Arrott plots. The respective values are summarized in Tab.~\ref{Mag_Data}. In Fig.~\ref{MT}(b) the inverse magnetic
susceptibility $1/\chi = H/M$ as well as the fit functions following the Curie-Weiss{} law $\chi-\chi_0=C/(T-\theta_{\rm CW})$ for $T>T_{\rm C}$ are
plotted. Here, $\chi_0$ accounts for a sum of the temperature-independent diamagnetic, Van-Vleck, and Pauli susceptibilities. The linear dependence
demonstrates the paramagnetic behavior in a broad temperature range, and the extracted values for the Curie-Weiss{} temperature $\theta_{\rm CW}$ are
again in good agreement with Ref.~\cite{Hor2010}. From the Curie constant $C = xN_{\rm A}\mu_{\rm eff}^2/3k_{\rm B}$, the effective magnetic moment
$\mu_{\rm eff}$ per Mn ion can be calculated. Here, $N_{\rm A}$ and $k_{\rm B}$ are the Avogadro number and the Boltzmann constant. The obtained
values lie in between $5.5\mu_{\rm B}$ and $5.8\mu_{\rm B}$ which is only slightly below the expected value for the Mn$^{2+}$ ion calculated from
$\mu_{\rm cal}=g\sqrt{S(S+1)}=5.9\mu_{\rm B}$ with the spin $S=\frac{5}{2}$ and the g-factor $g \approx 2$ (see Sect.~\ref{ESR}). Only for $x=0.04$
the value $\mu_{\rm eff}=4.3\mu_{\rm B}$ is somewhat lower, what could be a hint of a lower actual value of $x$. Altogether the estimates of
$\mu_{\rm eff}$ point to a Mn$^{2+}$ valence state in agreement with the XAS results in Ref.~\cite{Vobornik2011}. The 2+ valency of Mn is further
proved by the ESR data (see Sect.~\ref{ESR}).
\begin{table}
    \begin{tabular}{l r r r r r r}
    \hline \hline\noalign{\smallskip}
    $x$ & $T_{\rm C}$\,{\cite{Hor2010}} & $\theta_{\rm CW}$\,{\cite{Hor2010}} & $\theta_{\rm CW}$ & $\mu_{\rm eff}$ & $\mu_0 \frac{\mathrm{d}\Delta H}{\mathrm{d}T}$  \\\noalign{\smallskip}
        & (K) & (K) & (K) & ($\mu_{\rm B}$) & ($T/K$)\\

    \hline\noalign{\smallskip}
    0.01 & -   & -0.7 & -0.7 & 5.5 & -    \\

    \hline\noalign{\smallskip}
    0.04 & 9   & 11   & 9    & 4.3 & 1.21 \\

    \hline\noalign{\smallskip}
    0.07 & -   & -    & 10   & 5.5 & 1.19 \\

    \hline\noalign{\smallskip}
    0.09 & 12  & 13   & 11   & 5.8 & 0.89 \\

    \hline \hline
    \end{tabular}
\caption{Magnetic properties of \BTM\, single crystals. Values measured by \citet{Hor2010} are added for comparison.}
\label{Mag_Data}
\end{table}

\subsection{\label{Trans}Electrical Transport}

\begin{figure}
\includegraphics[width=1\columnwidth]{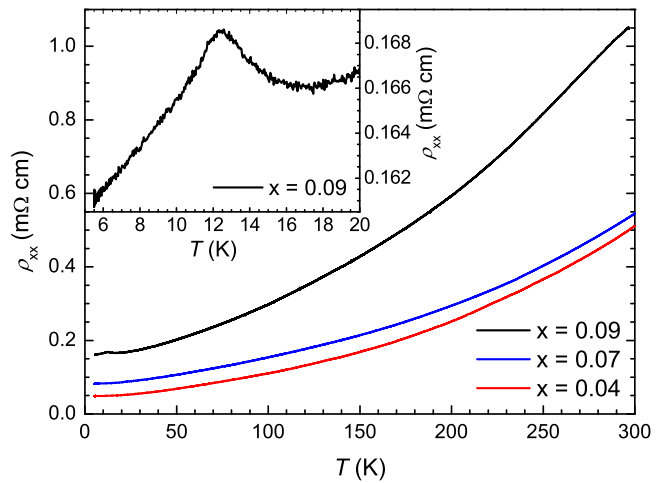}
\caption{Temperature dependence of the resistivity parallel to the charge current $\rho_{\rm xx}(T)$ of \BTM\ for various doping concentrations $x$. The
inset shows the enlarged low temperature regime for $x=0.09$.} \label{ResT}
\end{figure}
\begin{figure}
\includegraphics[width=0.92\columnwidth]{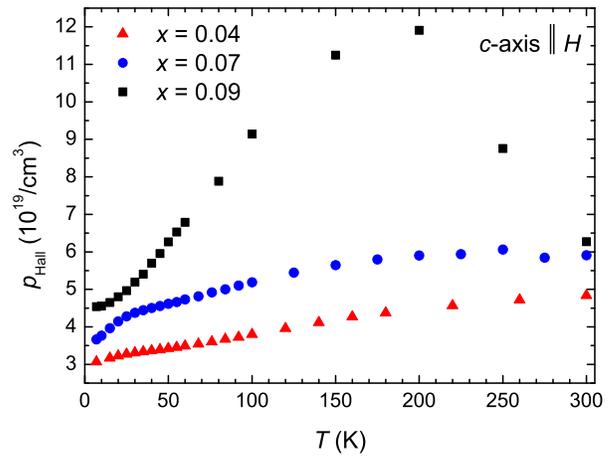}
\caption{Temperature dependence of the charge-carrier density $p_{\rm Hall}(T)$ of \BTM\ for various $x$. The values were obtained from the Hall
resistivity in the high-field region $\mu_0H \geq 11$\,T. (see the text)} \label{Hall_2}
\end{figure}

To obtain insights into the electronic properties, the temperature dependence of the resistivity $\rho_{\rm xx}(T)$ was measured. Furthermore, Hall
resistivity $\rho_{\rm xy}(T,H)$  measurements were performed to determine the charge-carrier density of the samples to be investigated by ESR.
\\
The resistivity (Fig. \ref{ResT}) shows a metallic temperature dependence as expected for a heavily doped semiconductor. Consistently with previous
measurements on crystals from the same batch \cite{Hor2010}, the resistivity increases with doping concentration. This could be explained by an increased number of
scattering centers with higher Mn doping. As can be seen in the inset of Fig.~\ref{ResT}, the resistivity of the highest doped sample shows an
anomaly around 12\,K associated with the ferromagnetic phase transition. In this case, the increased scattering is connected with an enhancement of
magnetic fluctuations near the phase transition, and it was similarly observed before in comparable systems \cite{Dyck2002}.
\\
\begin{figure}
\includegraphics[width=0.92\columnwidth]{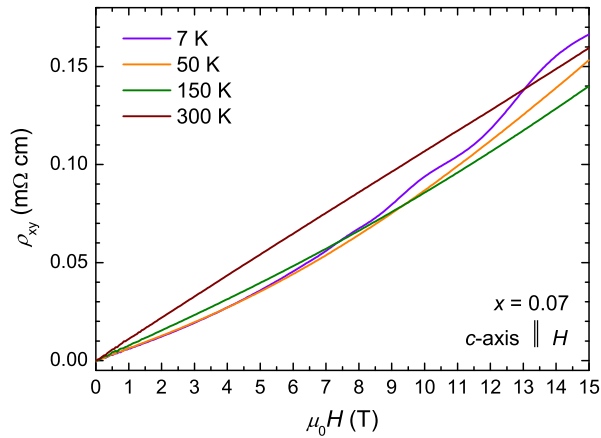}
\caption{Magnetic field dependence of the Hall-resistivity $\rho_{\rm xy}(H)$ of \BTM\ for $x=0.07$ at various temperatures. Measurements were performed
in the $ab$ basal plane, the external field $H$ was applied normal to this plane ($\| c$ axis).} \label{Hall}
\end{figure}
The Hall measurements are anti-symmetrized with respect to the sign of the magnetic field $H$ ($\rho_{\rm xy}=\frac{1}{2}[\rho_{\rm xy}^{\rm
H>0}-\rho_{\rm xy}^{\rm H<0}]$) to exclude the influence of the contact geometry. As exemplarily shown in Fig. \ref{Hall} for $x=0.07$, the Hall
resistivity shows a linear field dependence $\rho_{\rm xy}=R_{\rm H}\mu_0H$ at $T=300$\,K. With the Hall constant $R_{\rm H}$, the charge-carrier
density $p_{\rm Hall}=\frac{1}{eR_{\rm H}}$ is calculated from the high-field ($\mu_0H \geq 11$\,T) values of $\rho_{\rm xy}$ because in this range
the magnetization is saturated \cite{Hor2010}. As can be seen in Fig.~\ref{Hall_2}, the values of $p_{\rm Hall}$ for all three samples are positive
and in the range $10^{19}\, \mathrm{cm}^{-3} \lesssim p_{\rm Hall} \lesssim 10^{20}\, \mathrm{cm}^{-3}$. $p_{\rm Hall}$ increases with $x$, thus
confirming the acceptor behavior of Mn$^{2+}$ dopants.  The charge-carrier concentration is lower than the doping concentration ($c_{\rm Mn} \approx
2 - 5 \cdot 10^{20}\, \mathrm{cm}^{-3}$) showing that the crystals are partially electrically compensated.
\\
With decreasing temperature the field dependence of $\rho_{\rm xy}$ becomes nonlinear, especially for low fields (Fig.~\ref{Hall}). A similar behavior was
observed in \BT\, by \citet{Rischau2013} giving two possible explanations: the anisotropy of the Fermi surface and the occupation of a second valence
band. The anisotropy of the Fermi surface was described with a six-ellipsoidal model by \citet{Koehler1976}, where it was also stated that for $p >
4\cdot 10^{-18}\, \mathrm{cm}^{-3}$ a second valence band is occupied. A further signature of a second valence band is the temperature dependence of
$p_{\rm Hall}$ \cite{Jaworski2009,Kulbachinskii1999}. It becomes more pronounced with increasing $x$, which is equivalent to a stronger shift of the Fermi
level.
\\
In addition to the nonlinearity at low fields, Shubnikov-de Haas (SdH) oscillations are visible for $x=0.07$ at high fields at a low temperature
$T=7$\,K (Fig.~\ref{Hall}).
The frequency of the oscillation $F$ is proportional to the cross-section of the Fermi surface $S_{\rm c}$ perpendicular to the external field $H$ ($H
\parallel c$):
\begin{equation}
\frac{1}{F}=\Delta\left(\frac{1}{B}\right)=\frac{2 \pi e}{\hbar S_{\rm c}}.
\end{equation}
$F$ can be easily determined from the period of the oscillations $\Delta$ when plotting the derivative of $\rho$ over $B^{-1}$ (not shown). Following the
six-ellipsoidal model \cite{Koehler1976,Kulbachinskii2012}, an anisotropy factor $\eta=1.515$ can be introduced to calculate the volume of the
ellipsoid $V$ from $S_{\rm c}$. Taking also into account a factor of six for the degeneracy of the levels, the charge-carrier density can be estimated as:
\begin{equation}
p_{\rm SdH}=6 \frac{2V}{(2\pi)^3}\approx 6 \frac{2}{(2\pi)^3} \frac{4}{3\sqrt{\pi}}  \eta S_{\rm c}^{\frac{3}{2}}.
\end{equation}
The observed frequency $F=34.6$\,T yields $p_{\rm SdH}=1.1\cdot10^{19}$\,cm$^{-3}$. This is in the same order of magnitude but somewhat smaller than the
value extracted from the Hall resistivity $p_{\rm H}=3.7\cdot10^{19}$\,cm$^{-3}$. Similar differences have been observed for \BT\,
\cite{Koehler1976,Jaworski2009,Jo2014} and \ST\, \cite{Kulbachinskii1999} and prove the contribution of a second valence band to $p_{\rm Hall}$. A second
SdH frequency was not observed, probably due to a smaller effective mass that enters the amplitude of the oscillation.

\subsection{\label{ESR}Electron Spin Resonance}
\begin{figure}
\includegraphics[width=0.9\columnwidth]{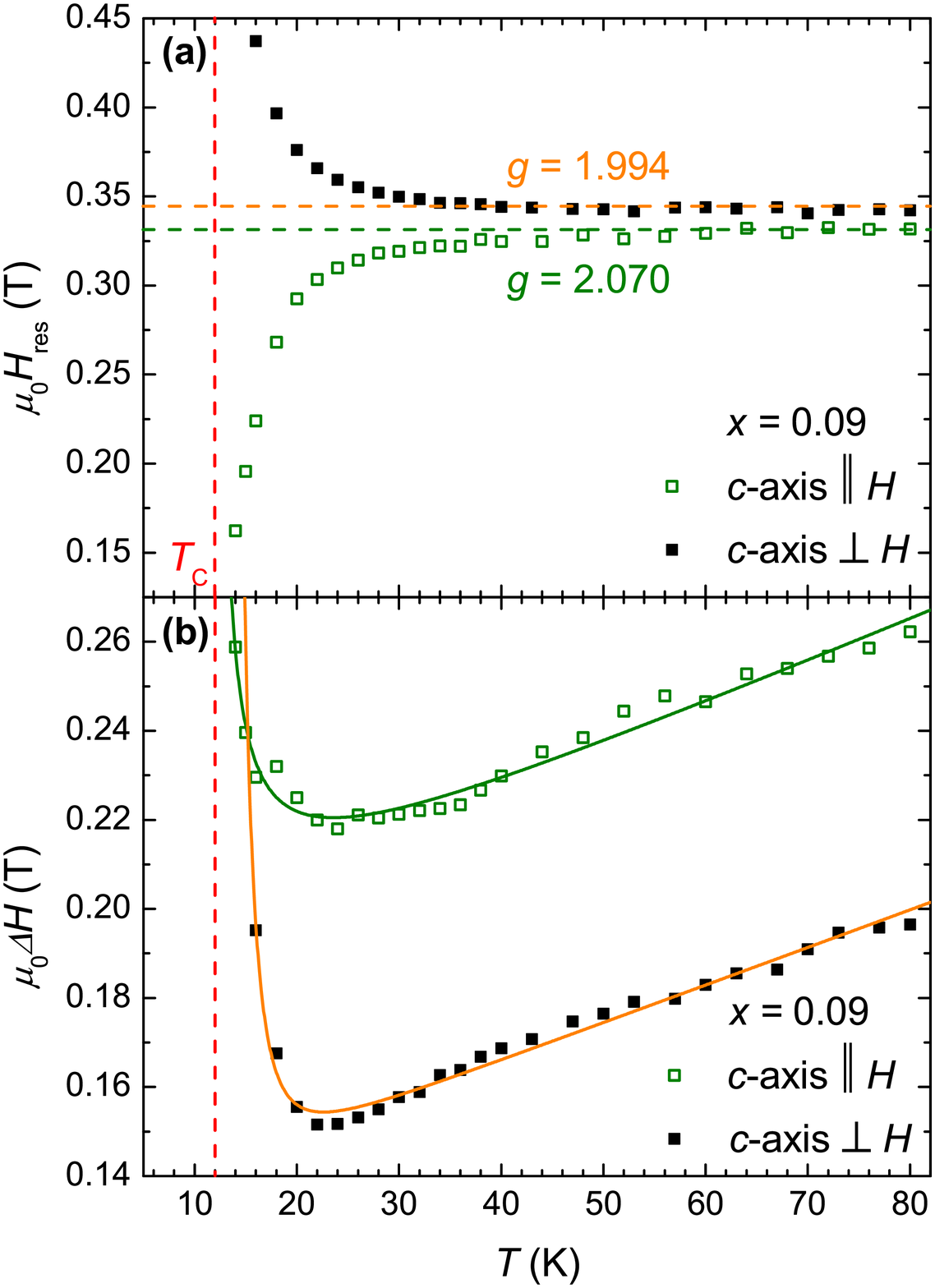}
\caption{Temperature dependence of the resonance field $H_{\rm res}(T)$ (a)  and  the linewidth $\Delta H(T)$ (b) of \BTM\ with $x=0.09$. The external
magnetic field $H$ is applied parallel and perpendicular to the crystallographic $c$ axis. The vertical dashed line marks the transition temperature
$T_{\rm C}$. The horizontal dashed lines in (a) denote the averaged values of the resonance field in the range $T=50-80$\,K with the corresponding values
of $g$. The solid lines in (b) are fits according to Eq.~\eqref{eq:linewidth_T}.} \label{ESR_Tdep}
\end{figure}

\begin{figure}
\includegraphics[width=0.9\columnwidth]{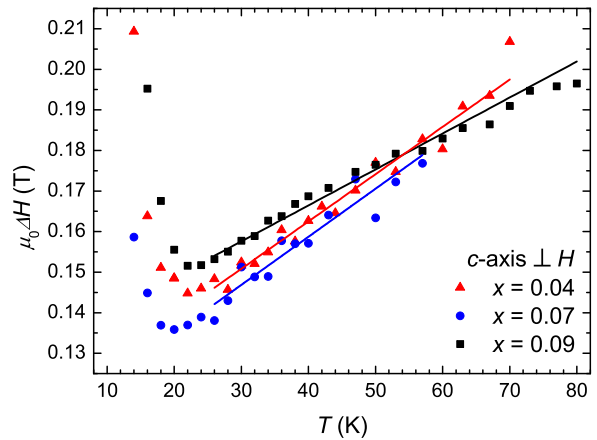}
\caption{Temperature dependence of the linewidth $\Delta H(T)$ of \BTM\ for various doping concentrations $x$. The external magnetic field $H$ was
applied perpendicular to the crystallographic $c$ axis. The solid lines are linear fits for $T\geq 25$\,K.} \label{dHT}
\end{figure}

ESR measurements at the X-band frequency $\nu \simeq 9.57$\,GHz were performed for five single crystals of \BTM\ with different doping concentrations
$x=0.01, 0.02, 0.04, 0.07, 0.09$. A single resonance line with a pronounced angular and temperature dependence was observed. The intensity of the
signal increases with $x$, so that the resonance signal can be undoubtedly assigned to the Mn dopants. The $g$-factor has a value $g\approx 2$ (for
$T\geq 50$\,K) very close to the free-electron value $g = 2.0023$. Such $g$-value is very typical for ESR of the Mn$^{2+}$ ions with $S=\frac{5}{2}$
and $L=0$ \cite{AbragamBleaney} thus confirming the 2+ valency of the Mn dopants. For $x\leq 0.02$ the intensity is weak so that the signal can only
be observed in a narrow temperature range and for the orientation of the magnetic field close to the $ab$ basal plane. Therefore, systematic studies
are presented for the samples with $x\geq 0.04$ with a particular focus on the sample with the highest doping ($x=0.09$).
\\
\begin{figure}
\includegraphics[width=0.94\columnwidth]{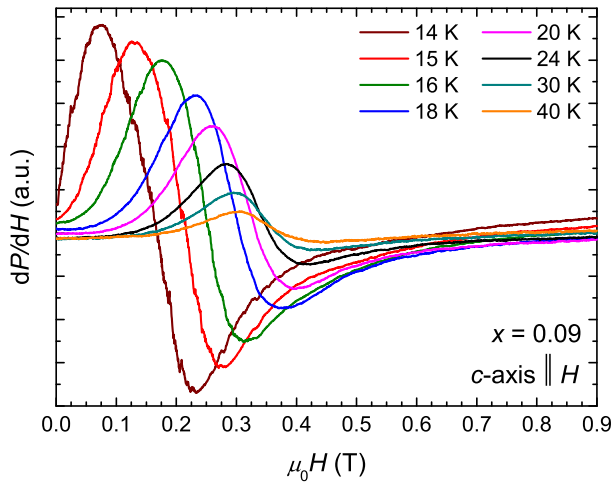}
\caption{ESR spectra of the \BTM\ sample with the highest doping level $x=0.09$  for various temperatures. The external magnetic field $H$ was
applied parallel to the crystallographic $c$ axis.} \label{ESRspec}
\end{figure}
A series of typical spectra (absorption derivatives $\mathrm{d}P(H)/\mathrm{d}H$) for different temperatures is plotted in Fig.~\ref{ESRspec}. The
ESR signal of all studied samples is always a single line. Its shape is asymmetric with a ratio of the positive and negative peaks of the absorption
derivative equal to 2.5. The signals can be fitted well with a Dysonian line profile, which is a mixture of the Lorentzian absorption and dispersion
derivatives yielding an accurate determination of the resonance field and linewidth (Fig.~\ref{ESR_Tdep}). A Dysonian ESR lineshape is typically
observed for localized magnetic moments in a metallic host, and it is due to a finite penetration depth of the microwaves in a bulk metallic sample
\cite{Barnes1981}. Similar lineshapes were also observed in Gd doped single crystals of \BT\, \cite{elKholdi1994,Isber1995} and \BS\,
\cite{Gratens1997}. Remarkably, no fine or hyperfine structure of the spectrum typical for ESR of isolated Mn$^{2+}$ ions \cite{AbragamBleaney} is
observed. A collapse of the spectrum into a single line with the Lorentzian (Dysonian) shape gives evidence for magnetic exchange interaction between
the Mn spins that averages and narrows the fine and the hyperfine structure and also reduces inhomogeneous line broadening originated from the
dipole-dipole interactions (exchange narrowing effect \cite{VanVleck48,Anderson53,Barnes1981}).
\\
As can be seen in Fig.~\ref{ESR_Tdep}(a), the resonance field $H_{\rm res}$ is practically constant at high temperatures. The dotted lines in
Fig.~\ref{ESR_Tdep}(a) represent the average for both orientations of the magnetic field in the range $T=50-80$\,K, and they correspond to g-factors of
$g=1.994$ for $H\parallel c$ and $g=2.070$ for $H\perp c$. A slight anisotropy of the $g$-factor arises due to a small admixture of the high-energy
multiplet with nonzero orbital momentum to the ground state spin-only multiplet of Mn$^{2+}$ \cite{AbragamBleaney}. In addition, in conducting
systems a covalent bonding between localized moments and conducting bands can lead to a shift of the $g$-factor \citep{Barnes1981}. The degree of
covalent bonding could depend on the orientation of the magnetic orbitals resulting in an anisotropy of $g$. For the samples with $x=0.04$ and $x=0.07$,
nearly identical values for $H_{\rm res}$ with the same temperature dependence were observed (not shown).
\\
Below $T\approx 40$\,K the resonance field develops a remarkable temperature dependence for both field orientations [Fig.~\ref{ESR_Tdep}(a)]. For
$H\parallel c$ axis, $H_{\rm res}$ shifts to lower fields while it increases for the perpendicular orientation. The shifts increase rapidly when
approaching the transition temperature $T_{\rm C}$. In a ferromagnet, the shift of the magnetic resonance signal from its paramagnetic position
$H_{\rm res}^{\rm param} = (h/g\mu_{\rm B})\nu$ determined by the $g$-factor is due to the shape anisotropy as well as the magnetocrystalline
anisotropy of the easy-axis or easy-plane type \cite{Turov65}. Magnetization measurements below $T_{\rm C}$ by \citet{Hor2010} have revealed that
\BTM\ is an easy-axis ferromagnet with the magnetic easy-axis parallel to the crystallographic $c$ axis. From the $M(H)$ curves, the saturation
magnetization $\mu_0 M_{\rm s} \approx 0.01$\,T and the anisotropy field $\mu_0 H_{\rm A} \approx 1$\,T were also determined \cite{Hor2010}. In this case
the ferromagnetic resonance signal should shift with respect to $H_{\rm res}^{\rm param}$ to smaller fields for $H\parallel c$ and to higher fields
for $H \perp c$ \cite{Turov65}. On the other hand, the shape anisotropy of the plate-like single crystalline sample would cause an opposite effect
\cite{Kittel48}. Comparing the magnetocrystalline anisotropy constant $K_{\rm u} = (\mu_0/2)H_{\rm A} M_{\rm s} \approx 4$\,kJ$/$m$^3$ with the shape
anisotropy constant $K_{\rm d}$, it becomes clear that even for the limiting case of a thin layer ($K_{\rm d} = - (\mu_0/2)(4\pi/3) M_{\rm s}^2
\approx - 0.2$\,kJ$/$m$^3$), the intrinsic magnetocrystalline part of the anisotropy is dominating. In the present work no signals could be detected
below $T_{\rm C}$ because ferromagnetic resonance is out of the frequency range of the used X-band spectrometer ($\nu = 9.6$\,GHz) \cite{commentFMR}
due to the opening of the magnetic anisotropy gap for the ESR excitation, $\nu_{\rm MA} = (g\mu_0 \mu_{\rm B}/h) H_{\rm A} \approx 28$\,GHz. However,
the dependence of the shifts of the ESR signal on the field orientation in the paramagnetic state above $T_{\rm C}$ strongly resembles the shifts
expected in the ferromagnetically ordered state with the dominant easy-axis type ($\parallel c$ axis) magnetocrystalline anisotropy. Thus, it is
reasonable to conclude that these shifts originate from the short-range ferromagnetic correlations between the Mn spins, which could be static on the
fast timescale of an ESR measurement of the order $\nu^{-1}\sim 100$\,ps. Considering the temperature dependence of $H_{\rm res}$, these correlations
persist up to temperatures much higher than $T_{\rm C}$, which is a typical signature of low-dimensional magnets \cite{Benner1990}. Indeed, a
low-dimensional character of the spin-spin interactions can be expected in the two-dimensional crystallographic structure of \BTM. This observation
is supported by the theoretical findings of \citet{Vergniory2014}, where the exchange interaction in the atomic layers was found
to be stronger than that in between the layers.\\
The temperature dependence of the Mn$^{2+}$ ESR linewidth also reveals two distinct regimes as depicted in Fig.~\ref{ESR_Tdep}(b) for two
orientations of the external field. For $T\gtrsim 25$\,K the linewidth increases linearly with $T$, which is an expected type of dependence for ESR
of paramagnetic local moments in a metal. It has been described by Korringa \cite{Korringa1950,Barnes1981} as a result of the relaxation of localized
spins due to their coupling to the spins of conduction electrons with energies within the $k_{\rm B}T$ region around the Fermi energy $E_{\rm F}$. Thus, the
so-called Korringa relaxation rate scales linearly with temperature. The temperature dependent part of the linewidth due to the Korringa relaxation
reads \cite{Korringa1950,Barnes1981}:
\begin{equation}
\Delta H_{\rm Kor}(T)=\frac{2H_{\rm res}}{\hbar \nu} \left[D(E_{\rm F}) J_{\rm cd}\right]^2 k_{\rm B}T. \label{Korringa}
\end{equation}
Here $D(E_{\rm F})$ is the density of states at the Fermi level and $J_{\rm cd}$ is the exchange integral between the conduction electrons (or holes) and the
$d$ electrons of the Mn impurities. The observation of the linear Korringa dependence is proof of the coupling between localized $d$ electrons with
the conducting holes.
\\
The temperature dependence of $\Delta H$ for three investigated samples with $x$ equal to 0.04, 0.07 and 0.09 is compared in Fig.~\ref{dHT}. The
linear fit in the temperature range above 25\,K yields the values of the slope $b=\mu_0(\mathrm{d}\Delta H /\mathrm{d}T)$ equal to $(1.21 \pm 0.17)$\,mT/K,
$(1.19 \pm 0.12)$\,mT/K and $(0.89 \pm 0.10)$\,mT/K for $x=0.04,\, 0.07$ and 0.09 respectively. The Korringa slope $b$ changes only slightly with the
doping concentration $x$. Hence, the system shows a non-bottlenecked behavior, i.e., the coupled subsystems of the Mn localized spins, conducting
holes, and the lattice are in equilibrium, and the spin angular momentum of Mn transferred to the reservoir of charge carriers has enough time to
decay to the lattice \cite{Barnes1981}.
For $T\lesssim 25$\,K the linewidth increases rapidly when approaching the transition temperature $T_{\rm C}$. This critical broadening can be ascribed to
the slowing down of the spin fluctuations resulting from the enhancement of magnetic correlations in the short-range spin-correlated regime above
$T_{\rm C}$. Taking into account a temperature-independent term $\Delta H_0$ and the linear-in-$T$ Korringa term [Eq.~(\ref{Korringa})], the $\Delta H(T)$
dependence in the entire temperature range can be described as:
\begin{equation}
\Delta H(T) = \Delta H_0 + \Delta H_{\rm Kor}(T) + c \left( \frac{|T-T_{\rm C}|}{T_{\rm C}}\right) ^{-\rm \beta} .
\label{eq:linewidth_T}
\end{equation}
Here, the last term represents the broadening near $T_{\rm C}$ with the critical exponent $\beta$ and the temperature-independent prefactor $c$. As can be
seen in Fig.~\ref{ESR_Tdep}(b), the experimental data are described well by Eq.~\eqref{eq:linewidth_T}. The determination of $\beta$ is quite
sensitive to the value of $T_{\rm C}$ resulting in an appreciable uncertainty. With $T_{\rm C}=12$\,K, the critical exponent $\beta \approx 0.9 \pm 0.3$ is
obtained for $H
\parallel c$. This is close to the theoretically expected value for a three-dimensional Heisenberg magnet $\beta^{\rm 3D}_{\rm Heisenberg} =1$
\cite{Benner1990}. Following the arguments of \citet{Benner1990} in the critical regime, the moments can behave Ising-like due to the internal
fields. Comparing the determined exponent with the Ising critical exponents, one can see that it lies in between the three- and two-dimensional values
$\beta^{\rm 3D}_{\rm Ising} =0.6$ and $\beta^{\rm 2D}_{\rm Ising} =1.3$, respectively, in accord with the presumed low-dimensional character of the Mn-Mn magnetic
correlations in \BTM. For $H \perp c$, a steeper slope is observed with $\beta \approx 2.8 \pm 0.6$ since here the field is oriented perpendicular to
the easy-axis. For $x=0.04$ and $x=0.07$, similar values for the critical exponent were determined.
\begin{figure}
\includegraphics[width=0.9\columnwidth]{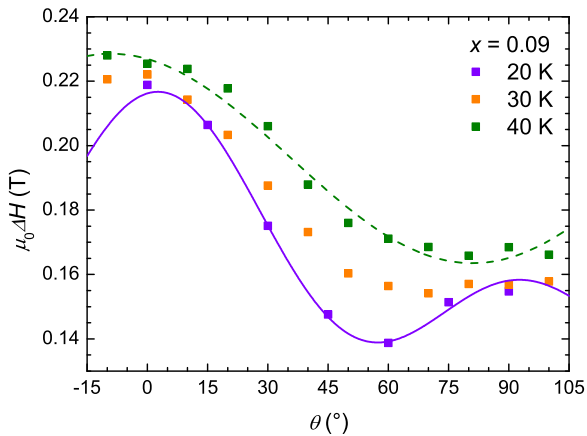}
\caption{Angular dependence of the linewidth $\Delta H(\theta)$ of \BTM\ for $x=0.09$. $\theta$ is the angle between the external magnetic field $H$
and the crystallographic $c$ axis. The lines correspond to fits of the data (see the text). } \label{Aniso}
\end{figure}
Finally, the angular dependence of the linewidth is shown in Fig.~\ref{Aniso} for three different temperatures. For $T=20$\,K, it follows that $\Delta H
\propto (3 \cos ^2 \theta -1)^2$, which has a characteristic minimum at the so-called ``magic" angle $55\,^\circ$ (solid line), a dependency that
results from the anisotropic dipole-dipole interaction. This observation is characteristic for two-dimensional systems in the critical regime
\cite{Benner1990} where the slowing down of the spin dynamics makes the exchange narrowing of the ESR signal less effective and  enhances the
dipole-dipole contribution to the linewidth. For higher temperatures, this effect is less dominant. For $T=40$\,K, the angular dependence resembles
$\Delta H \propto (\cos ^2 \theta +1)$ as marked by the dashed line. This behavior was also observed in \BSM\, \citep{Bardeleben2013}. It is
characteristic for three-dimensional systems with dominant exchange narrowing, and it signifies the occurrence of the uncorrelated truly paramagnetic
regime in \BTM\ above $T\approx 40$\,K. The $\Delta H(\theta)$ dependence at $T = 30$\,K appears to be intermediate between these two types.
\section{\label{Disc}Electronic properties and exchange integrals}
As follows from Eq.~\eqref{Korringa}, the Korringa slope $b=\mu_0\mathrm{d}\Delta H /\mathrm{d}T$ depends on the density of states at the Fermi
level, $D(E_{\rm F})$, and the exchange integral between conducting holes and localized Mn moments, $J_{\rm cd}$. It is possible to estimate $J_{\rm cd}$ by
calculating $D(E_{\rm F})$ from the charge-carrier density $p_{\rm Hall}$ with $D(E_{\rm F})=4 \frac{m^*}{h^2}(3\pi^2p_{\rm Hall})^{1/3}$ using the simple model of a
three-dimensional electron gas. From our transport measurements in Sect.~\ref{Trans}, the mobile charge carriers in \BTM\ are holes with a charge-carrier density in the range $10^{19}\, \mathrm{cm}^{-3} \lesssim p_{\rm Hall} \lesssim  10^{20}\, \mathrm{cm}^{-3}$. The effective mass of the valence
band $m^*=0.35m_{\rm e}$ was determined by \citet{Koehler1976}. The values of the slope $b$ are similar for the three samples ($x=0.04, 0.07$ and 0.09), resulting in similar values for the exchange constant in the range $J_{\rm cd} \approx  0.5 -0.7$\,eV. This is a typical scale
for magnetically doped semiconductors ($J_{\rm cd}\lesssim 1$\,eV) \cite{Matsukura1998}. For V doped \ST\, a larger value was observed with
$J_{\rm cd}=5.3$\,eV. \citet{Li2014} had predicted $J_{\rm pd}=1.4$\,eV for the exchange between the Mn $d$ orbitals and the Te $p$ orbitals that form the
valence band. The above estimation is oversimplified since it ignores the anisotropy of the Fermi surface (see Ref.~\cite{Koehler1976}) and the
probable existence of a second partially occupied valence band, as is suggested by the analysis of our transport measurements. In a two-band model, the Korringa relaxation rate should have two contributions, making it difficult to calculate $J_{\rm cd}$ without further input. Nevertheless, the
simple one-band model can be used to clarify the consistency of our data. For the RKKY interaction, the exchange integral $J_{\rm ij}$ between two
localized moments at sites $i$ and $j$ is defined as:
\begin{equation}
J_{\rm ij} =  \frac{3m^* V^2}{4\pi h^2} J_{\rm cd}^2 \frac{\sin(2k_{\rm F}r) -2k_{\rm F}r\cos(2k_{\rm F}r) }{r^4}
\label{eq:TI_RKKY}
\end{equation}
Here, $V$ is the volume of the unit cell, $k_{\rm F}$ is the Fermi vector, and $r$ is the distance between the moments at sites $i$ and $j$ that can be calculated
from the Mn concentration. Using again the model of the 3D electron gas and the estimated values for $J_{\rm cd}$, one obtains $|J_{\rm ij}| \approx 2-3$\,meV
with $J_{\rm ij} < 0$ (ferromagnetic). This agrees with theoretical predictions for Mn in \BT\, from \citet{Vergniory2014} ($|J_{\rm ij}| \lesssim
2$\,meV) and \citet{Henk2012} ($|J_{\rm ij}|\approx 2$ - 3\,meV). Besides, the Curie-Weiss temperature can be calculated \cite{Matsukura1998} with
$\theta_{\rm CW} = \left[2S(S+1)/3k_{\rm B}\right] \cdot \left[xz\cdot |J_{\rm ij}|/5\right]$ using $J_{\rm ij}$ from above, $S=5/2$, and the number of nearest
neighbors $z=6$. The results are values in the range $\theta_{\rm CW}\approx 3$ - 9\,K that increase with the doping concentration $x$. Considering the
crudeness of the approximation these values are in fair agreement with the measured values $\theta_{\rm CW}\approx 9$ - 11\,K (Table.~\ref{Mag_Data}). Therefore, the exchange coupling is
presumably dominated by one band instead of two. This would imply different values of $D(E_{\rm F})$, $J_{\rm cd}$ or $m^*$ for the
proposed two bands.
\\
Furthermore, the coupling of the Mn moments and the mobile holes results in an anisotropy of $H_{\rm res}$ in the paramagnetic regime at $T \geq 40$\,K
(Fig.~\ref{ESR_Tdep}). The additional contribution to the anisotropy due to the crystal-field effect originates from the anisotropy of the band
structure \cite{Barnes1981}, which in this case is most likely determined by the layered crystal structure.
\section{\label{Concl}Conclusion}
The spin dynamics and magnetic interactions between Mn dopants in \BTM\, high-quality single crystals have been investigated using electron spin
resonance spectroscopy and complementary measurements of the static magnetization and transport properties. Magnetization measurements show the
existence of a ferromagnetic phase below a transition temperature $T_{\rm C} \approx 12$\,K for $x\geq 0.04$. The electrical transport measurements reveal
$p$-type conductivity with a charge-carrier density in the range $10^{19}\, \mathrm{cm}^{-3} \lesssim p_{\rm Hall} \lesssim 10^{20}\, \mathrm{cm}^{-3}$.
A detailed study of the temperature and orientational dependences of the ESR signal of the Mn$^{2+}$ dopants reveals two distinct regimes: (i) a
paramagnetic uncorrelated regime above $T\sim 20 - 40$\,K which is characterized by the relaxation of Mn spins on the mobile charge carriers
(Korringa relaxation); and (ii) a critical spin-correlated regime below $T\sim 20 - 40$\,K which is characterized by anisotropic shifts of the position
of the Mn$^{2+}$ ESR signal and its critical broadening by approaching the ordering temperature $T_{\rm C}$. The observed Korringa behavior of the ESR
linewidth gives evidence that the localized Mn moments are exchange-coupled to the mobile holes. A combined qualitative discussion of the ESR and
transport data enables the conclusion that the charge carriers mediate ferromagnetic interaction between the Mn dopants via the RKKY mechanism with a
coupling strength $|J_{\rm ij}| \approx 2-3$\,meV, thus confirming the theories of \citet{Li2014} and \citet{Vergniory2014}, where RKKY-type interactions are predicted. The critical behavior of the
ESR signal at lower temperatures indicates a gradual development of the  Mn spin-spin correlations with the easy-axis type ($\parallel c$ axis)
magnetic anisotropy. These correlations are visible in the ESR data at temperatures far above $T_{\rm C}$, suggesting a low-dimensional character for the
magnetic exchange between the Mn dopants that can be ascribed to the layered crystal structure. Our experimental findings appear useful for a deeper
understanding of magnetic interactions in magnetically doped topological insulators.

\section{Acknowledgments}

This work was supported in part by the Deutsche Forschungsgemeinschaft (DFG) through project KA 1694/8-1 and the collaborative Research Center SFB 1143, and at Princeton by the NSF MRSEC program,
grant DMR-1420541.

\bibliography{Bi2Te3Mn}

\end{document}